\documentclass[aps,prl,twocolumn,groupedaddress,nofootinbib,notitlepage,showpacs,floatfix]{revtex4-1}

\usepackage{graphicx,graphics,epsfig,subfigure,times,bm,bbm,amssymb,amsmath,amsfonts,amsthm,mathrsfs,MnSymbol}
\usepackage[matrix,frame,arrow]{xypic}
\usepackage[pdfstartview=FitH]{hyperref}
\usepackage{pifont}
\hypersetup{
  colorlinks=true,       	
    linkcolor=red,          	
   citecolor=magenta,        
    filecolor=magenta,      	
    urlcolor=cyan,           	
    runcolor=cyan
}
\usepackage[pdftex]{color}
\usepackage{hypernat}
\usepackage{braket}
\usepackage{enumerate}
\usepackage[normalem]{ulem}

\usepackage{subfigure}
\usepackage{overpic}

\usepackage{multirow}
\newcommand{\be}{\begin{equation}}
\newcommand{\ee}{\end{equation}}
\newcommand{\ba}{\begin{eqnarray}}
\newcommand{\ea}{\end{eqnarray}}

\newcommand{\ignore}[1]{}

\begin{document}


\title{Superfluid qubit systems with ring shaped optical lattices}

\author{Luigi Amico}
\affiliation{ CNR-MATIS-IMM $\&$ 
Dipartimento di Fisica e Astronomia,   Via S. Sofia 64, 95127 Catania, Italy}
\affiliation{ Centre for Quantum Technologies, National University of
Singapore, 3 Science Drive 2, Singapore 117543}
\affiliation{Institute of Advanced Studies, Nanyang Technological
University, 1 Nanyang Walk, Singapore 637616}

\author{Davit Aghamalyan}
\affiliation{ Centre for Quantum Technologies, National University of
Singapore, 3 Science Drive 2, Singapore 117543}

\author{ P. Auksztol, H. Crepaz, R. Dumke} 
\affiliation{Centre for Quantum Technologies, National University of
Singapore, 3 Science Drive 2, Singapore 117543}
\affiliation{Division of Physics and Applied Physics, Nanyang Technological University, 21 Nanyang Link,
Singapore 637371}
\author{ L.-C. Kwek}
\affiliation{ Centre for Quantum Technologies, National University of
Singapore, 3 Science Drive 2, Singapore 117543}
\affiliation{National Institute of
Education and Institute of Advanced Studies, Nanyang Technological
University, 1 Nanyang Walk, Singapore 637616}


\begin{abstract}
We study an experimentally feasible qubit system employing neutral atomic currents.
Our system is based on bosonic cold atoms trapped in ring-shaped optical lattice potentials.
The lattice makes the system strictly one dimensional and it provides the infrastructure to realize a  tunable ring-ring interaction.
Our implementation combines the low decoherence rates of of neutral cold atoms systems, overcoming single site addressing, with the robustness of topologically protected solid state Josephson flux qubits. Characteristic fluctuations in the magnetic fields affecting Josephson junction based flux qubits are expected to be minimized employing neutral atoms as flux carriers. By breaking the Galilean invariance we demonstrate how atomic currents through the lattice provide a implementation of a qubit. This is realized either by artificially creating a phase slip in a single ring, or by tunnel coupling of two homogeneous ring lattices. The single qubit infrastructure is experimentally investigated with tailored optical potentials.
Indeed, we have experimentally realized scaled ring-lattice potentials that could host, in principle,
$n\sim 10$ of such  ring-qubits, arranged in a stack configuration, along the laser beam propagation axis.
 An experimentally viable scheme of the two-ring-qubit is discussed, as well.
 Based on our analysis, we provide protocols to initialize, address, and read-out the qubit. \vspace{5mm}
\end{abstract}

\maketitle




\section*{Introduction}
\label{intro}
A qubit is a two state quantum system that can be coherently manipulated, coupled to its neighbours, and measured.
Several qubit physical  implementations have been proposed in the last decade, all of them presenting specific virtues and bottlenecks at different levels\cite{superconducting,coldatoms,rydberg,iontraps,NMR,qdots}.
In neutral cold atoms proposals the qubit is encoded into well isolated internal atomic states.
This allows long coherence times, precise state readout and, in principle, scalable quantum registers.
However, individual qubit (atom) addressing is a delicate point\cite{addressing1, addressing2}.
Qubits based on Josephson junctions allow fast gate operations and make use of the precision reached by lithography techniques\cite{Martinis}.
The decoherence, however, is fast in these systems and it is experimentally challenging to reduce it.
For charge qubits the main problem arises from dephasing due to background charges in the substrate;
flux qubits are insensitive to the latter decoherence source, but are influenced by magnetic flux fluctuations due to impaired spins proximal to the device\cite{superconducting}.

Here we aim at combining the advantages  of cold atom and Josephson junction based implementations.
The basic idea is  to use  the  persistent currents flowing through ring shaped optical lattices\cite{luigi,hallwood_homogeneous,rey_homogeneous,rey_non-homo,hallwood_delta} to  realize a cold atom analogue of the superconducting flux qubit (see~\cite{rot1,rot2,synth1,berry,luigi} for the different schemes that can be applied to induce  persistent currents).
Recently, superpositions of persistent currents  have been  thoroughly investigated\cite{rey_non-homo,hallwood_delta}.

\section*{Results}
\label{results}
In  this paper we demonstrate how  persistent currents flowing in a ring shaped optical lattice can provide a physical implementation of  a qubit\cite{luigi,rey_non-homo,hallwood_delta}.  The lattice potential plays an important role in our approach.  Indeed, it makes strictly one dimensional the atoms' dynamics. Further it provides the means for precise control of the confinement and facilitates the qubit-qubit interaction.
In our system we break the Galilean invariance. For a single ring this is realized by creating a localized 'defect' barrier along a homogeneous lattice\cite{slm}. Additionally we prove that a qubit can be achieved with two homogeneous interacting rings arranged vertically on top of each other. In such a system the Galilean invariance is broken along the direction transverse to the two rings. For this scheme we analyse the real time dynamics and time-of-flight density distributions.  Based on our analysis, we provide viable protocols to initialize, address, and read-out the qubits.  {Indeed, we have experimentally realized scaled ring-lattice potentials that could host, in principle,
$n\sim 10$ ring-qubits, arranged in a stack configuration, along the laser beam propagation axis}.

 \vspace{0.2cm}

\subsection*{\bf Single-ring-qubit: Breaking the Galilean invariance on the single ring with a site defect.}
\label{single_ring}
\begin{figure}
\centering
\rotatebox{0}{\includegraphics [width=88mm]{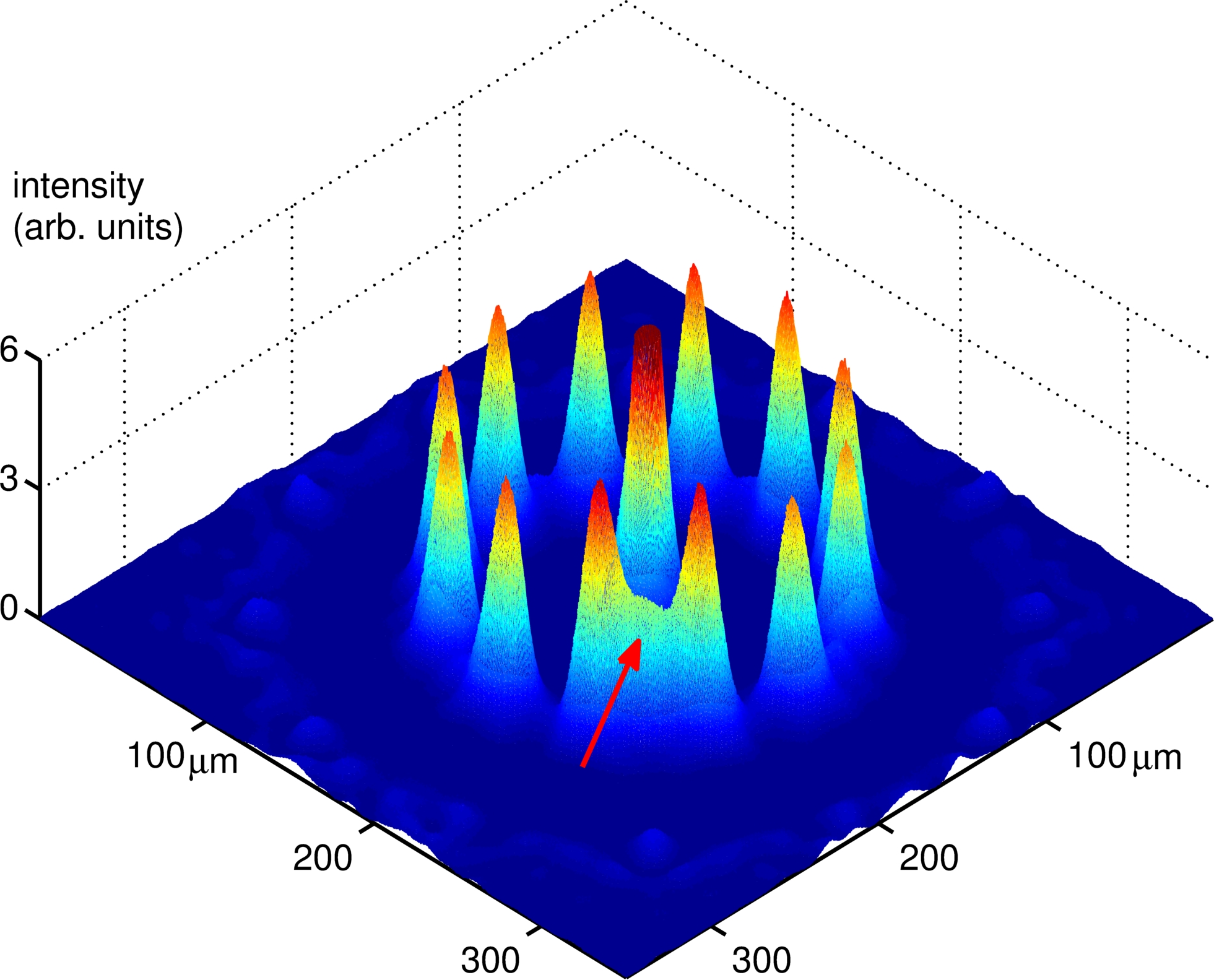} }
\caption{{\bf\sffamily Experimental realization of a ring-lattice potential with an adjustable weak link (red arrow).}
Measured intensity distribution with an azimuthal lattice spacing of 28 $\mathsf{\mu m}$ and a ring radius of 88 $\mathsf{\mu m}$ (see Methods section). The centre peak is the residual zero-order diffraction.  The effective dynamics
of a condensate in such a system is governed by the qubit potential as discussed in Eq.(2). The size of the structure is scalable and a lower limit is imposed by the diffraction limit of the focusing optics (see Methods section). Several rings can be arranged in a stack, along the propagation axis of the laser beam (shown in Fig.\ref{lattice_axial}).}
\label{exp-onering}
\end{figure}

We consider bosonic atoms loaded in a ring-shaped potential with identical wells, but with a  dimple located at the site $N-1$ (see Fig.\ref{exp-onering}), and pierced by a 'magnetic flux' $\Phi$. The system is described by the Bose-Hubbard Hamiltonian

\begin{equation}
H_{BH}=  \frac{U}{2}\sum_{i=0}^{N-1}n_{i}(n_{i}-1) -\sum_{i=0}^{N-1} t_i (e^{i\Phi/N} a_{i}^{\dag}a_{i+1}+h.c.)
\label{BH}
\end{equation}
where  $a_{i}$'s are bosonic operators for atoms trapped in the  ring and $n_{i}\doteq a_{i}^{\dag}a_{i}$.
The parameters $t_i$   describe the tunnelling  between the wells along the ring. Since the wells are all identical but one, $t_i=t,  \forall i=0\dots N-2$ and $t_{N-1}=t'$. Finally,  U describes the  s-wave scattering interaction \cite{BRUDER}.
The 'magnetic flux' is $\Phi=\int_{x_{i}}^{x_{i+1}}\emph{\textbf{A(z)}}d\emph{\textbf{z}}$, where $\emph{\textbf{A(z)}}$ is the effective  vector potential.  The effect of the dimple is to induce a phase slip at the site $N-1$.
We assume that the density of superfluid  is large enough to neglect  the fluctuation of the number of atoms in each well. In this regime we can assume that the system  dynamics is characterized by the phases of the superfluid order parameter $\phi_i$'s, described by the
quantum phase model \cite{Fazio_Rev} with Josephson coupling    $J_i \sim \langle n \rangle t_i $ ($\langle n \rangle$ is the average number of bosons in each well). The magnetic flux $\Phi$ can be gauged away everywhere but at the site $(N-1)$-th \cite{schulz-shastry}.  Accordingly, the  phase difference along nearest neighbour sites can be considered small in the 'bulk' and the harmonic approximation can be  applied.
The partition function can be written as a path integral: $Z=\int {\cal D}[\phi] e^{-S[\phi]}$, where the  $S[\phi]$ is the Euclidean action.
Adapting from  the approach pursued by Rastelli et al. \cite{rastelli}, all  the phases $\phi_i$ except $\theta\doteq \phi_{N-1}-\phi_0$   can be integrated out (the integrals are Gaussian). The effective action reads
\begin{align}
S_{eff} = \int_0^\beta d \tau & \left [ {\frac {1}{2 U}} \dot{\theta}^2 + \frac{J}{2(N-1)} (\theta-\Phi)^2 -J' \cos(\theta) \right ]  \nonumber  \\
  - & \frac{J}{2 U(N-1)} \int d \tau d \tau' \theta(\tau)G(\tau-\tau') \theta(\tau')
\label{effective_single}
\end{align}
with the potential $\displaystyle{U(\theta)\doteq \frac{J}{N-1} (\theta-\Phi)^2 -J' \cos(\theta)}  $.   For large $(N-1)J'/J$  and moderate $N$, $U(\theta)$ defines a two-level system.  The degeneracy point is $\Phi=\pi$: The two states are provided by the symmetric and antisymmetric combination of counter-circulating currents corresponding to the two minima of $U(\theta)$.
We observe that breaking the Galilean invariance of the system provides an independent parameter $J'$  facilitating the control  of the potential landscape.  The interaction between $\theta$ and the (harmonic)  bulk degrees of freedom provides the non local term with  $G(\tau)=\sum_{l=0}^\infty Y(\omega_l) e^{i\omega_l \tau}$, $\omega_l$ being Matsubara frequencies and $\displaystyle{Y(\omega_l)=  \omega_l^2 \sum_{k=1}^{(N-2)/2} \frac{1+\cos[2 \pi k /(N-1)}{2 JU(1-\cos[2 \pi k /(N-1)])+\omega_l^2}}$.
The external bath vanishes in the thermodynamic limit and the effective action reduces to the Caldeira-Leggett one \cite{rastelli}.
Finally it is worth noting that the case of a single junction needs a specific approach but it can be demonstrated consistent with Eq.(\ref{effective_single}).
 \medskip

\subsection*{\bf Two-rings-qubit: Breaking the Galilean invariance with two homogeneous coupled rings.}
\begin{figure}
\centering
{\includegraphics [width=88mm]{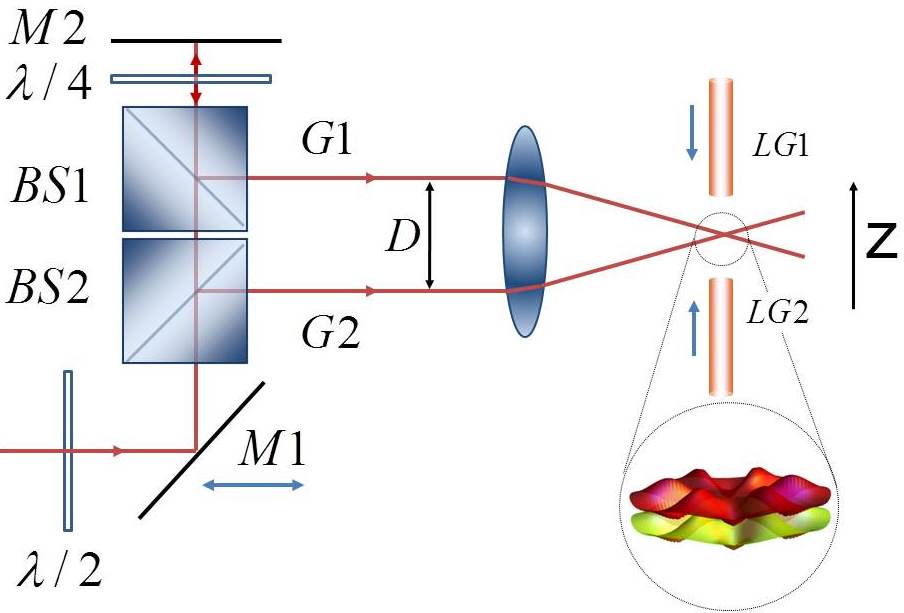} }
\caption{{\bf\sffamily Setup for the ring-ring coupling.}
Two parallel Gaussian laser beams (G1,G2) are produced by a combination of two polarizing beamsplitter (BS1, BS2).  The beam separation $D$ can be controlled by moving mirror $M_1$. Both beams  pass through a lens and interfere to form a lattice in z-direction.
The distance between the lattice planes is a function of $1/D$ \cite{Li} which can be varied. The resulting one dimensional lattice is combined with vertical beams (LG1, LG2) providing horizontal confinement for trapped atoms (See the Methods section).
The inset shows the ring lattice potentials separated by  $d=\lambda_{1} {f}/{D}$ \cite{Li}.
The ring-ring separation is adjustable by varying the distance $D$. Such an arrangement provides an effective two-level system that can be exploited as a qubit (See text)..}
\label{setup-tworings}
\end{figure}

We consider bosonic atoms loaded in two coupled  identical {\it homogeneous} rings Fig.\ref{setup-tworings}. We will prove that such a system effectively provides a qubit-dynamics (alternatively to the one-ring qubit implementation discussed above).
The system is described by the Bose-Hubbard ladder:   $H=H_{BH}^{(a)}+H_{BH}^{(b)}+H_{int}$, where $H_{BH}^{(a,b)}$ are the Hamiltonians as in Eq.(\ref{BH}) for  the bosons in the rings $a$ and $b$ respectively, and 
\label{hom_coupled_ring}
\begin{equation}
\label{two_rings_int}
 H_{int}=-g\sum_{i=1}^{N}(a_{i}^{\dag}b_{i}+b_{i}^{\dag}a_{i}).
 \end{equation}
 We  observe that  along each ring the phase slips  imply  twisted boundary conditions and therefore they can be localized to a specific site,  say the $N-1$-th.
Following a similar procedure as employed above, the effective action reads
\begin{align}
S_{eff}= \int_0^\beta & d \tau \left [ {\frac {1}{2 U}}  \sum_{\alpha=a,b} \dot{\theta_\alpha}^2 + U(\theta_a,\theta_b) \right ]    \\
- \frac{J}{2 U(N-1)} \sum_{\alpha=a,b} & \int d \tau d \tau' \theta_\alpha(\tau)G_\alpha(\tau-\tau') \theta_\alpha(\tau') \nonumber
\label{effective-two}
\end{align}
where each $G_\alpha(\tau)$ is  given by the expression found above for the case of a  single ring. In this case the phase dynamics is provided by the potential
\begin{align}
U(\theta_a,\theta_b)\doteq \sum_{\alpha=a,b} & \left [ \frac{J}{2(N-1)} (\theta_\alpha-\Phi_\alpha)^2 -J \cos(\theta_\alpha) \right ] \nonumber  \\
& - \tilde{J}  \cos[\theta_a-\theta_b-{\frac{N-2}{N}}(\Phi_a-\Phi_b)]  \; .
\end{align}
with $\tilde{J}=\langle n \rangle g $\cite{davit}.  We observe that, for large $N$, the potential  $U(\theta_a,\theta_b)$ provides  that  effective phase dynamics of Josephson junctions flux qubits realized by Mooji {\it et al.}  (large $N$'s corresponds  to large geometrical inductance of flux qubit devices) \cite{fluxJJ}. In there, the landscape was thoroughly analysed.  The qubit is made with superpositions of  the two states $|\theta_1 \rangle $ and $|\theta_2 \rangle $  corresponding to the minima of  $U(\theta_a,\theta_b)$. The degeneracy point is achieved by  $\Phi_b-\Phi_a=\pi$. We comment that the ratio  $\tilde{J}/J $ controls the relative size of the energy barriers between  minima intra- and minima inter-'unit cells' of the $(\theta_a,\theta_b)$ phase space, and therefore is important for designing the qubit. In our system $\tilde{J}/J $ can be fine tuned with the scheme shown in Fig.\ref{setup-tworings}.

{ Having established that the two tunnel-coupled homogeneous rings, indeed, define a two level system,
we now study its real-time dynamics}. We will show that the density of the condensate in the two rings  can display characteristic oscillations in time.

\begin{figure}
\centering
\includegraphics [width=88mm]{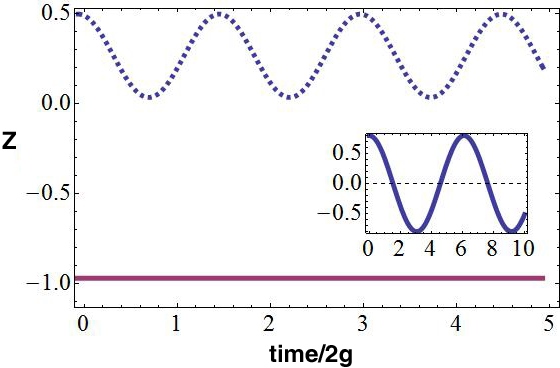}
\caption{{\bf\sffamily Population imbalance  in  two coupled rings.}
We focused on the case  $\Delta \gg \lambda \rho$. For moderate   $z_0$, oscillations are obtained, with $\omega=2g\left \{\sqrt{1+\Delta^2}+ \lambda\rho {(z_0\Delta- \sqrt{1-z_0^2})(2\Delta^2-1)}/{[2(1+\Delta^2)^{3/2}]}\right \}$ corresponding to macroscopic quantum self trapping (blue dashed line). The dynamics can be visualized with the help of  the mechanical system provided by a  rotator of length $\sqrt{1-z(s)^2}$,  driven by the external force $\Delta$.  The constant solution $z(s)=const$ corresponds  to vanishing pendulum length (magenta solid line). For  $\Delta  = 0$ (inset), the dynamics is characterized by  Rabi oscillation with $\omega_0\doteq 2g (1+\lambda \rho \sqrt{1-z_0}/2)< \omega$.  Here $\lambda \rho=0.1$  and $\Delta=4$ implying that $\omega\approx 4 \omega_0$.}
\label{fig:imbalance}
\end{figure}

We make use of the  mean field approximation  to analyse the (real time) dynamics of the Bose-Hubbard ladder Eqs.(\ref{BH}), (\ref{two_rings_int}) (assuming that each ring is in a deep superfluid phase).
Accordingly   Gross-Pitaevskii equations are found for the quantities depending on the time $s$.
$ \varphi_{a,i}(s)=\langle a_{i}(s)\rangle$ and $\varphi_{b,i}(s)=\langle b_{i}(s)\rangle$.
Assuming that  $\theta_{\alpha}\doteq \varphi_{\alpha,i+1}- \varphi_{\alpha,i}$ in each ring is site-independent, we obtain
\begin{align}
 \frac{\partial z }{\partial \tilde{s}}  &=-\sqrt{1-z^{2}}\sin{\Theta}   \nonumber \\
 \frac{\partial \Theta }{\partial \tilde{s}} &=\Delta +\lambda \rho z+\frac{z}{\sqrt {1-z^{2}}}\cos{\Theta}
\label{dynamics}
\end{align}
where  $z=(N_{b}-N_{a})/(N_{a}+N_{b})$ is the normalized  imbalance between the populations $N_a$ and $N_b$ of the two rings, $\Theta=\theta_{a}-\theta_{b}$ and $ \tilde{s}\doteq 2gs$ ($\hbar=1$). The parameters  are $\Delta =\left [\mu_a -\mu_b + t(\cos{{\frac{\Phi_{a}}{N}}}-\cos{{\frac{\Phi_{b}}{N}}})\right ]/g$,   $\lambda=U/(2g)$, and $\rho=(N_{a}+N_{b})/N$ is the total bosonic density (we included the chemical potential $\mu_\alpha$).
Eqs.(\ref{dynamics}) can be solved analytically in terms of elliptic functions\cite{davit,Smerzi}.
Accordingly, the   dynamics displays distinct regimes (oscillating or exponential) as function of the elliptic modulus $k$, depending  in turn on  $\Delta$, $\lambda$, and    on the initial population imbalance $z(0)\doteq z_0$.
Here we consider the dynamics at $\lambda \rho \ll \Delta$, {\it i.e.} small $U/g$ (the analysis of the solutions of the Eqs.(\ref{dynamics}) in different regimes will be presented elsewhere).  The results are summarized  in  Fig.{\ref{fig:imbalance}}.   We comment that, comparing with $\Delta = 0$,  the oscillations  do not  average to zero (therefore yielding a macroscopic quantum self trapping phenomenon\cite{Smerzi}) and they are faster.
The pattern of the circulating currents along the two coupled rings can be read out through the analysis of the time-of-flight density.
As customarily, the spatial density distribution in the far field corresponds to the distribution in the momentum space at the time  when the  confinement potential is turned off:

\begin{align}
\label{eq:density}
& \rho(\textbf{k})=\frac {|w(k_x, k_y, k_z)|^{2}}{N} \sum_{i=0}^{N-1}\sum_{j=0}^{N-1}\sum_{q\in \{ 2\pi n/N\}} \\
& \left [ \cos{ [\textbf{k}_{\parallel}\cdot \textbf{x}_{\parallel} +(q+\frac{\Phi_{a}}{N}) (\phi_i-\phi_j)] } \langle a_{q}^{\dag}a_{q} \rangle +   \nonumber  \right .\\
& \left. \cos{ [\textbf{k}_{\parallel}\cdot \textbf{x}_{\parallel}+(q+\frac{\Phi_{b}}{N}) (\phi_i-\phi_j)] } \langle b_{q}^{\dag}b_{q}  \rangle +   \nonumber  \right .\\
 & \left . 2\cos{ [\textbf{k}_{\parallel}\cdot \textbf{x}_{\parallel} +k_{z}D+(q+\frac{\Phi_{a}}{N}) \phi_i-(q+\frac{\Phi_{b}}{N})\phi_j) ]}
\langle a_{q}^{\dag}b_{q}  \rangle \right ] \nonumber
\end{align}

where $w({k_x,k_y,k_z})$ are  Wannier functions (that we considered identical for the two rings),
$\textbf{k}_{\parallel}\cdot \textbf{x}_{\parallel}\doteq k_{x}(x_{i}-x_{j})+ k_{y}(y_{i}-y_{j}) $,
$x_{i}=\cos{\phi_i}$,  $y_{i}=\sin{\phi_i}$ fix the positions of the ring wells in the three dimensional space, $\phi_i=2\pi i /N$ being lattice sites  along the rings;
the expectation values involving the Fourier transforms of operators $a_q\doteq1/\sqrt{N}\sum_l e^{-i \phi_i l }a_l$ and $b_q\doteq1/\sqrt{N}\sum_l e^{-i \phi_l q }b_l$  are obtained for $U/t=0$.
The density Eq.(\ref{eq:density}) is displayed in Fig.\ref{timeoflight}

\begin{figure}
\centering
 \begin{tabular}{cc}
{\includegraphics [width=88mm]{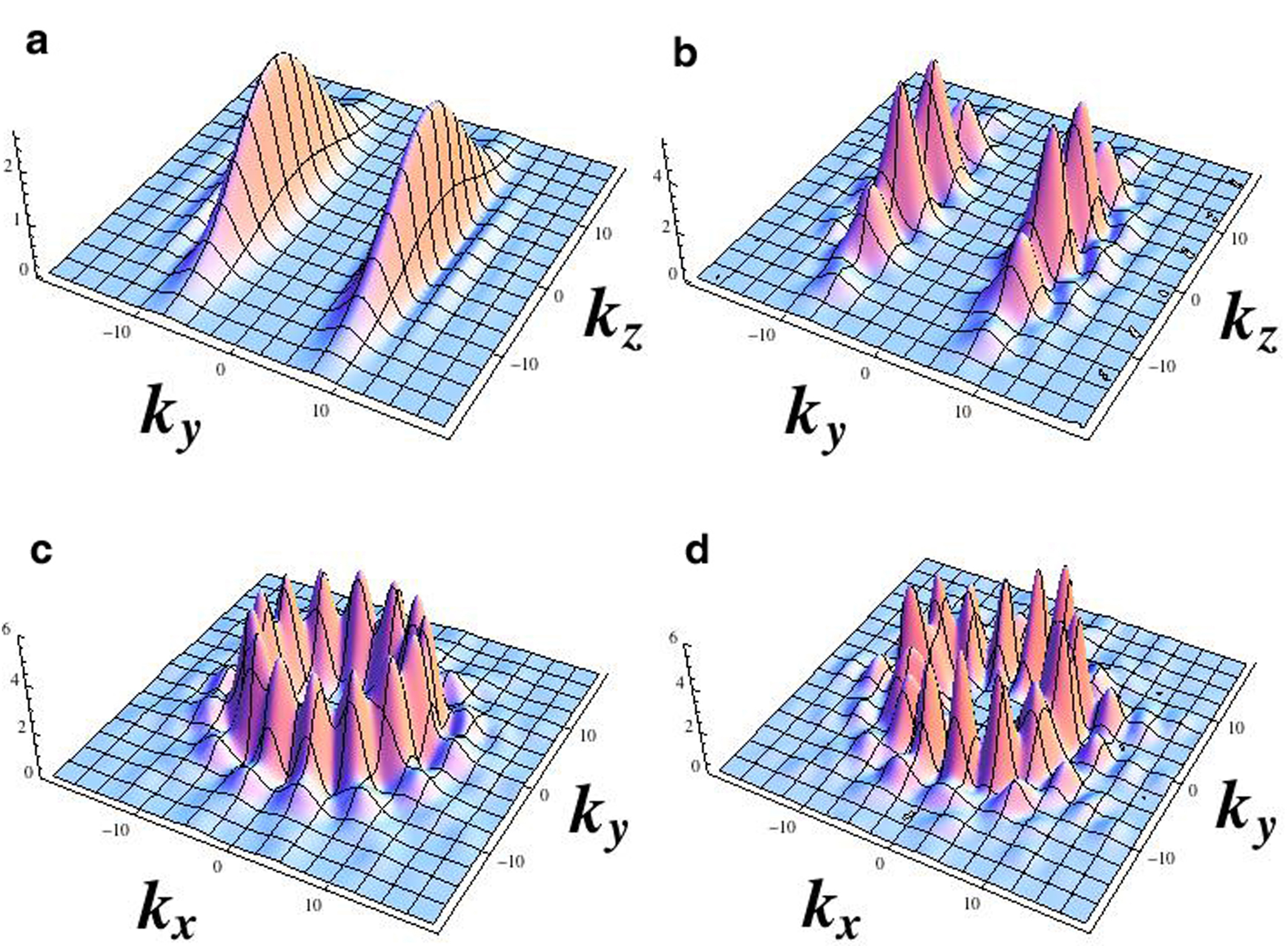}} 
 \end{tabular}
\caption{{\bf\sffamily Time-of-flight expansion for the two-coupled-rings-qubit.}
{\bf\sffamily a,c,} vanishing inter-ring tunnelling rate $g/t=0$. In {\bf\sffamily b,d,} $g/t=0.9$. In the $(k_x,k_y)$ plane the interference fringes with the ring symmetry are due to the momenta of the quantum degenerate gas; the inter-ring tunnelling suppresses the interference fringes. In the  $(k_y,k_z)$ plane, $g$ induces structured interference  fringes. The Eq. (\ref{eq:density}) is calculated for  the Bose-Hubbard ladder with  'fluxes' $\Phi_a$ and $\Phi_b$,  with  $U=0$, and at quantum degeneracy. Results are shown for $\Phi_a=80, \Phi_b=70$, $T=0.05 k_B$ and $N=14$ with filling fractions of $10$ bosons per site.}
\label{timeoflight}
\end{figure}

\section*{Discussion}
We  proposed  a construction of flux qubits with atomic neutral currents flowing in ring-shaped optical lattice potentials.
Persistent currents  had been experimentally observed in a narrow toroidal trap with a weak link\cite{phillips}. The effective action of the system  studied in\cite{russians} can provide a two level system.  In contrast with\cite{phillips,russians,Wright},  we emphasize how  we make explicit use of the lattice in our construction, both to confine the particles in the rings and to drive the ring-ring interaction. The qubits are realized by breaking the Galilean invariance of the system either by adding an additional barrier along a single ring lattice Eqs.(\ref{effective_single}),  or by tunnel coupling of two homogeneous rings, Eq.(\ref{effective-two}). The latter is  proposed to be realized  with the scheme in Fig.\ref{setup-tworings}.
We observe that a suitable variation  of such set-up can be exploited also  to create two qubit gates  (each qubit provided by Fig.\ref{exp-onering}); {alternatively, a route described in the Methods section can be pursued.} 

 The analysis of the real time dynamics of such system can be recast to a type of coupled Gross-Pitaevskii equations that are characteristic for  double well potentials, this providing  a further  proof that the system indeed defines a qubit. Accordingly, the basic phenomenology of the tunnel-coupled homogeneous rings is demonstrated to be characterized by macroscopic quantum self trapping.
Since  different flow states lead to characteristic density patterns in the far field, standard expansion of the condensate can be exploited to detect the different quantum states of the system (See Fig.\ref{timeoflight}). 

 Our work provides a feasible route to the implementation of a functional flux qubit based on persistent atomic currents. {{For an extensive discussion on the one and two qubit gates, please  refer to the Methods section.}}
The initialization of our qubit can be accomplished, for example, imparting rotation by exploiting  light induced torque from Laguerre-Gauss (LG) beams carrying optical angular momentum. A two-photon Raman transition between internal atomic states can then be used to transfer coherently $\hbar$ orbital angular momentum to the atoms.  With this method, transfer efficiencies of 90\% to the rotating state had been demonstrated\cite{phillips,greiner}. Owing to the coherent nature of the Raman process, superpositions of different angular momentum states can be prepared\cite{kapale}.  Measurements of the decay dynamics of a rotating condensate in an optical ring trap showed remarkable long lifetimes of the quantized flow states on the order of tens of seconds even for high angular momentum (l=10). Phase slips - the dominant decoherence mechanism - condensate fragmentation and collective excitations which would destroy the topologically protected quantum state are strongly suppressed below a critical flow velocity. Atom loss in the rotating condensate doesn't destroy the state but leads to a slow decrease in the robustness of the superfluid where phase slips become more likely\cite{zoran,perrin}. 

\begin{figure}
\centering
{\includegraphics [width=88mm]{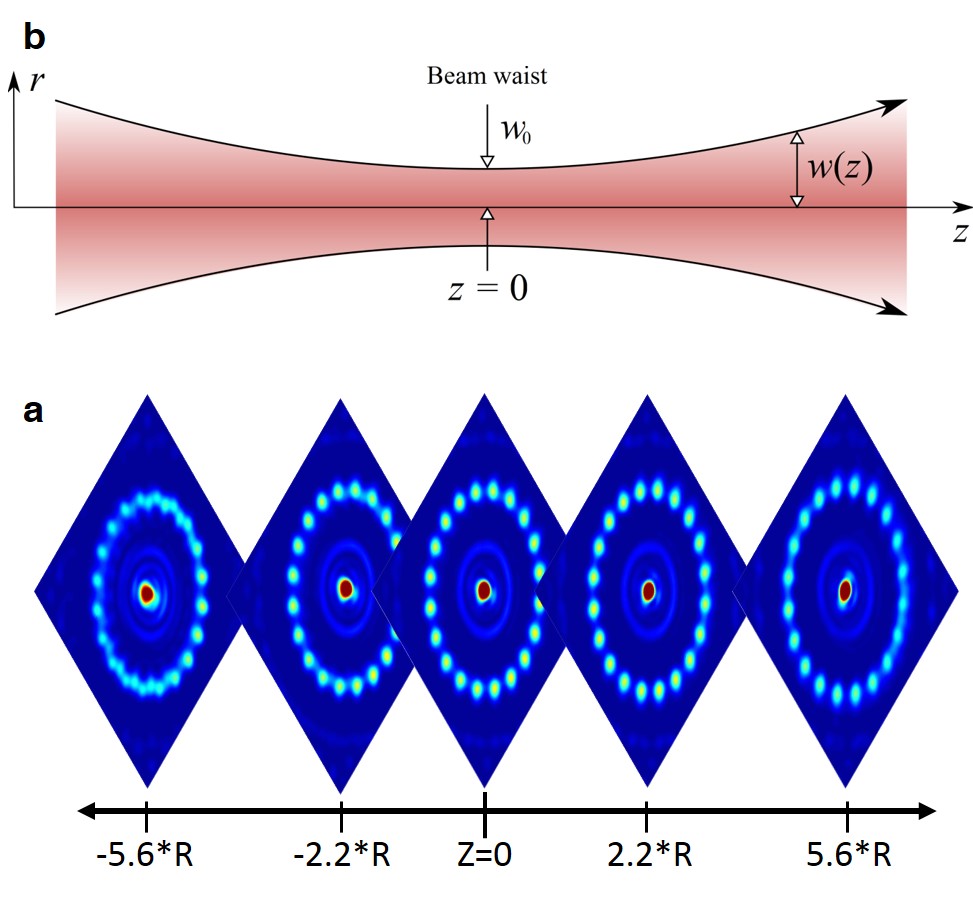} }
\caption{{\bf\sffamily Effect of an axial translation on the ring lattice potential.}
{{\bf\sffamily (a)} {Ring lattice intensity distribution measured at various positions along the beam propagation axis around the focal plane (Z=0). Note that the initial beam, phase modified by the SLM, is not Gaussian any more.} The optical potential remains undisturbed by a translation of 2.2 times the ring-lattice radius centred around the focal plane (Z=0). Here R designates the ring-lattice radius of 87.5 $\mathrm{\mu m}$. {\bf\sffamily (b)} This is in contrast to a Gaussian laser beam which exhibits a marked dependence on the axial shift from the focal plane where the beam waist $\omega(z)$ scales with $\sqrt{1+(z/z_{0})^{2}}$ and Rayleigh range $z_{0}$.}}
\label{lattice_axial}
\end{figure}

We comment that, because of the lattice confinement, the gap between the two levels of the qubit displays a favourable scaling with the number of atoms in the system (assuming that the temperature is low enough we can describe the system with Eq.(\ref{BH}))\cite{hallwood_delta,rey_non-homo,Hallwood_scaling}.
{Besides making the inter-ring dynamics strictly one dimensional, the lattice confinement provides  the route to  the inter-rings coupling. 
Indeed, the light intensity  results to be  modulated along the (nearly) cylindrical laser beam.  Analysing  our experimental configuration, we conclude that  it is feasible  to arrange $n\sim 10$  ring-qubits in  stacks configuration (as sketched in Fig.\ref{lattice_axial}) along the beam propagation axis.
To allow controlled tunnelling between neighbouring lattice along the stack, the distance between the ring potentials needs to be adjustable in the optical wavelength regime (the schematics in Fig.\ref{setup-tworings} can be employed). A trade-off between  high tunnelling rates (a necessity for fast gate operations)  and an efficient  read out and addressability  of individual stack sites, needs to be analysed.} Increasing the lattice stack separation  after the tunnelling interaction has occurred well above the diffraction limit while keeping the atoms confined, optical detection and addressing of individual rings becomes possible.
{
\\
This arrangement produces equal, adjustable ring-ring spacing between individual vertical lattice sites and can therefore not readily be used to couple two two-ring qubits to perform two-qubit quantum-gates. The SLM method, however, can be extended to produce two ring-lattices in the same horizontal plane, separated by a distance larger than the ring diameter. The separation between these two adjacent rings can then be programmatically adjusted by updating the kinoform to allow tunnelling by mode overlap\cite{anderson}. Combined with the adjustable vertical lattice (shown in Fig.\ref{setup-tworings}) this would allow, in principle, two-ring qubit stacks to be circumferential tunnel-coupled to form two-qubit gates.}

Read out of the angular momentum states can be accomplished experimentally with interference of different flow states (i.e. corresponding to a fragmented superfluid) which maps the phase winding into a density modulation that can be measured using time-of-flight imaging\cite{zoran}. In  the lower panel of Fig.\ref{timeoflight} it is shown that different flow states lead to characteristic density patterns in the far field.

We believe that our implementation combines the advantages of neutral cold atoms and solid state Josephson junction based flux qubits for applications in quantum simulation and computation. This promises to exploit the typically low decoherence rates of the cold atom systems, overcoming the single site addressing\cite{Tian}, and harness the full power of macroscopic quantum phenomena in topologically non trivial systems.  The characteristic fluctuations in the magnetic fields affecting Josephson junction based flux qubits are expected to be minimized employing neutral atoms as flux carriers.

\section*{Methods}
\label{methods}
{\small
\subsection*{\bf Experimental realization of the ring-lattice potential with weak link}
\label{exp_ring_lattice}
We created the optical potential with a liquid crystal on silicon spatial light modulator (PLUTO phase only SLM, Holoeye Photonics AG) which imprints a controlled phase onto a collimated laser beam from a 532 nm wavelength diode pumped solid state (DPSS) laser.
The SLM acts as a programmable phase array and modifies locally the phase of an incoming beam. Diffracted light from the computer generated phase hologram then forms the desired intensity pattern in the focal plane of an optical system (doublet lens, f=150 mm). The resulting intensity distribution is related to the phase distribution of the beam exiting the SLM by Fourier transform.
Calculation of the required SLM phase pattern (kinoform) has been carried out using an improved version of the Mixed-Region-Amplitude-Freedom (MRAF) algorithm\cite{slm,slm1} with angular spectrum propagator. This allows us to simulate numerically the wavefront propagation in the optical system without resorting to paraxial approximation. A region outside the desired ring lattice pattern (noise region) is dedicated to collect unwanted light contributions resulting from the MRAF algorithm's iterative optimization process. This can be seen in the measured intensity pattern in Fig.1 as concentric, periodic structures surrounding the ring-lattice and can be filtered out by an aperture.

{The ring-lattice potential shown in Fig.2 and Fig.5 can be readily scaled down from a radius of $\mathrm{\sim 90\; \mu m}$ to $\mathrm{5-10\; \mu m}$ by using a 50x microscope objective with NA=0.42 numerical aperture (Mitutoyo 50x NIR M-Plan APO) as the focusing optics for the SLM beam and with $\mathrm{\lambda_{2}=830\,nm}$ light, suitable for trapping Rubidium atoms.}  Accounting for the limited reflectivity and diffraction efficiency of the SLM, scattering into the noise region and losses in the optical system only about 5\% of the laser light contributes to the optical trapping potential. However this is not a limiting factor for small ring-lattice sizes in the tenth of micrometer range as discussed here where $\sim 50$ mW laser power is sufficient to produce well depths of several $\mathrm{E_{rec}}$. The generated structures are sufficiently smooth, with a measured intensity variation of 4.5\% rms, to sustain persistent flow-states\cite{phillips}.
The barrier height can be dynamically modified at a rate up to 50 ms per step, with an upper limit imposed by the frame update rate of the SLM LCD panel (60 Hz). \newline

\subsection*{\bf Setup for the adjustable ring-ring coupling}
To allow controlled tunnelling between neighbouring lattice stacks the distance between the ring potentials needs to be adjustable in the optical wavelength regime. Small distances allow high tunnelling rates, a necessity for fast gate operations.
This makes it  less efficient to read out and address individual stack sites, however. Increasing the lattice stack separation  after the tunnelling interaction has occurred well above the diffraction limit (${\sim}\lambda$) while keeping the atoms confined, optical detection and addressing of individual rings becomes possible.
Fig.\ref{setup-tworings} in the main text illustrates the experimental arrangement to produce two  adjustable $1d$ ring-lattices by intersecting two Gaussian beams (G1,G2) with wavelength $\lambda_{1}$.
The inset in Fig.\ref{setup-tworings} shows two vertically spaced ring lattice potential separated by $d=\lambda_{1} {f}/{D}$ \cite{Li}.
The ring-ring separation is controllable by changing the beam spacing $D$ between beams $G1$ and $G2$, allowing adjustment of the ring-ring tunnelling. \\
In an experimentally feasible arrangement using light from a Ti:Sa laser at $\mathrm{\lambda_{1} \approx 830\, nm,}$ with a beam separation adjustable between $\mathrm{D=10-40}$ mm and a lens focal length $f=75$ mm, the ring-ring separation can be varied from $\mathrm{d=1.5-6.2\,\mu m.}$ This compares to a inter-ring well spacing of $\mathrm{1.5\, \mu m}$ for a ring lattice with 20 lattice sites and ring radius of $\mathrm{5\, \mu m}$.
{
Taking advantage of a large ring-ring separation of $\mathrm{5\, \mu m}$ facilitates addressing of individual rings to generate different effective flux-states in a stack. Circulation can be created, for instance, with a pulsed pair of Raman beams where one of the Raman beams carry $\hbar$ orbital angular momentum. By Raman coupling the $|F=2,m_{F}=0\rangle$ and $|F=2,m_{F}=2\rangle$ Zeeman ground-states manifolds of $^{87}Rb$ and employing a magnetic gradient field along the vertical axis, the effective two-photon Raman detuning can be shifted out of resonance for atoms in rings other than the addressed one. The differential Zeeman energy shift between the two Raman ground states leads to a  magnetic field dependent shift $ \delta = \mu_{B} g_{F} \Delta m_{F} B $ of the two-photon Raman detuning. Here $\mu_{B}$ denotes the Bohr magneton, $g_{F}$ the Land\'e  $g$-factor, $\Delta m_{F}$ the difference between the magnetic spin-quantum numbers of the two Raman states and B the magnetic field strength. 
With a magnetic field gradient of 180 G/cm -- a typical value for magnetic traps in BEC experiments -- the two-photon Raman detuning of a ring which is 5 $\mu$m separated from the addressed one with $\delta =0 $ would be shifted by $\delta =$126 KHz. As was shown by Wright \textit{et al.}\cite{wright}, with appropriate choices of the magnitude, intensity ratio and detuning of the Raman beams, fractional population transfer between the 
$ |2,2 \rangle \leftrightarrow |2,0 \rangle $ states can be accurately controlled by varying the two-photon Raman detuning $\delta$ in a range of less than $200$ KHz. This was demonstrated for Raman beams with Gaussian beam profiles and hence no orbital angular momentum was transferred onto the atoms but it can, in principle, be adapted for a combination of Gaussian and Laguerre-Gaussian beams to generate atomic flux states.  
}\\
With a SLM arbitrary optical potentials can be produced in a controlled way only in a $2d$-plane -- the focal plane of the Fourier transform lens -- making it challenging to extend and up-scale this scheme to 3d trap arrangements.
The experiment, however, showed (see Fig.\ref{lattice_axial}) that axially the ring structure potential remains almost undisturbed by a translation along the beam propagation axis of $\mathrm{\Delta z\, =\, \pm 2.2\cdot R}$, where $R$ denotes the ring-lattice radius.  {The ring-lattice radius is only weakly affected by an axial shift along z and scales with $\Delta R/R=0.0097\cdot z,$ where $z$ is normalized to the ring-lattice radius.}
For larger axial shifts from the focal plane the quality of the optical potential diminishes gradually. {Based on our measurements this would allow implementation of ring-lattice stacks with more than $10$ rings in a vertical arrangement, assuming a stack separation comparable to the spacing between two adjacent lattice sites.
Propagation invariant beams may allow a potentially large number of rings to be vertically arranged\cite{propinvariant}.} \newline

\subsection*{\bf Tunnelling rate estimation for the two coupled ring lattices.}
The ring lattice potential shown in the inset in Fig. \ref{setup-tworings} can be written as
\begin{align*}
V_{latt}=4E_{0}^{2}(f_{pl}^{2} \cos{(k_{LG}z)}^2+\cos{(k_{G}z)}^2+ \\
2f_{pl} \cos{(k_{LG} z)}\cos{(k_{G} z)} \cos{(\phi l)}),
\end{align*}
where $f_{pl}$ are related to Laguerre functions\cite{luigi}.
Such a potential with $l$ lattice sites can be created directly by diffraction from a SLM or by superposition of two Laguerre-Gaussian beams with a positive and negative azimuthal index $\pm l$, respectively\cite{scottish}.
The  WKB estimate of the tunnelling rate gives
\begin{equation}
t_z=4\sqrt{\frac{1}{\hbar \sqrt{2m}}}\frac{V_0^{3/4}}{\sqrt{d}}e^{-\frac{\sqrt{2m V_0}}{\pi\hbar}d}
\end{equation}
where $d=\lambda f/D$ is the lattice spacing along z-direction.

\subsection*{\bf Demonstration of the one qubit and two qubit unitary gates}
The aim of this section is to show how the effective phase dynamics of optical ring-lattices with impurities serves the  construction of  one- and two-qubit gates - a necessity for universal quantum computation. Here, we adapt results which were obtained by Solenov and Mozyrsky\cite{Solenov1} for the case of homogeneous rings with impurities. It results  that a single ring optical lattice with impurity is described by the following effective Lagrangian (see Eq. (\ref{effective_single}) and Supplemental information):
\be
L={\frac {1}{2 U}}  \dot{\theta}^2+\frac{J}{N-1} (\theta-\Phi)^2 -J' \cos \theta
\ee
Then we introduce the canonical momentum P in a usual way:
\be
P=\frac{\partial L}{\partial {\dot{\theta}}}=\frac {1}{ U}\dot{\theta}
\ee
After performing a Legendre transformation we get the following Hamiltonian:
\be
H=J'(\frac {P^2}{2 \mu}-\frac{J}{J'(N-1)} (\theta-\Phi)^2 + \cos \theta) \; ,
\label{H_double}
\ee
where $\mu=J'/U$ is an effective mass of the collective particle.
The quantization is performed by the usual transformation $P\rightarrow -d/d\theta$.
For $\delta=\frac{J'(N-1)}{2J}>1$ the effective potential in (\ref{H_double}) can be reduced to a double well; for $\Phi=\pi$, the two lowest  levels of such double well are symmetric and antisymmetric superpositions of the states in the left and right wells respectively (See the Supplemental material). The effective Hamiltonian  can be written as 
\be
H\simeq\varepsilon\sigma_z
\ee
and the lowest two states are $|\psi_g\rangle=(0,1)^{T}$ and $|\psi_e\rangle=(1,0)^{T}.$ An estimate for the gap energy can be found employing the WKB approximation \cite{Griffits} 
\be
\varepsilon\simeq \frac{2 \sqrt{UJ'}}{\pi}\sqrt{ (1-\frac{1}{\delta})} e^{-12\sqrt{ J'/U} (1-1/\delta)^{3/2} },
\ee
where $\delta>1$.
From this formula we can see that  the limit of weak barrier and strong interactions is most favourable regime to obtain a finite gap between the two energy  levels of the double level potential \cite{rey_homogeneous,rey_homogeneous1,hallwood_delta,Hallwood_scaling}. We also note  that the gap energy splitting can be controlled by the height of the impurity barrier. \newline

\subsection*{\bf Single qubit gates}
For the realization of single-qubit rotations, we consider the system close to the symmetric double well configuration $\Phi\simeq\pi$. In the basis of the two level system discussed before the Hamiltonian takes the form:
\be
H\simeq\varepsilon\sigma_z+\frac{\Phi-\pi}{\delta}\langle \theta \rangle_{01}\sigma_x,
\label{sigma}
\ee
where $\langle \theta \rangle_{01}$ is the off-diagonal element of the phase-slip in the two-level system basis. It is easy to show that spin flip, Hadamard and phase gates can be realized by this Hamiltonian. For example, a phase gate can be realized by evolving the  state through the  unitary transformation $U_z(\beta)$ (tuning the second term of Eq.({\ref{sigma}) to zero by adjusting the imprinted flux)
\be
U_z(\beta)=exp(i\varepsilon \tau\sigma_z)=\begin{pmatrix}e^{i \varepsilon \tau} & 0\\
0 & e^{-i \varepsilon \tau} 
\end{pmatrix} \; .
\ee
After tuning the gap energy close to zero (adjusting the barrier height of the impurity), we can realize the following rotation
\be
U_x(\beta)=exp(i\alpha \tau \sigma_x)=\begin{pmatrix}\cos{\alpha} & i\sin{\alpha}\\
i\sin{\alpha} & \cos{\alpha}
\end{pmatrix}
\ee
where $\alpha=\frac{\Phi-\pi}{\delta}\langle \theta \rangle_{01}\tau$. When  $\alpha=\pi/2$  and   $\alpha=\pi/4$ the NOT and  Hadamard gates are  respectively realized. \newline

\subsection*{\bf Two-qubit coupling and gates}
The effective dynamics for two coupled qubits, each realized as single ring with localized impurity (as in Fig.\ref{exp-onering}), is governed by the Lagrangian
\ba
L&=&  \sum_{\alpha=a,b} {\frac {1}{2 U}}\dot{\theta_\alpha}^2  +\left [ \frac{J}{2(N-1)} (\theta_\alpha-\Phi_\alpha)^2 -J' \cos(\theta_\alpha) \right ] \nonumber  \\
 &-& \tilde{J''}  \cos[\theta_a-\theta_b-{\frac{N-2}{N}}(\Phi_a-\Phi_b)]
\ea
Where $J''$ is the Josephson tunnelling energy between two rings. When $\Phi_a=\Phi_b=\Phi$ and $J''\ll J'$ the last term reduces to $-J''\frac{(\theta_a-\theta_b)^2}{2}$  and the Lagrangian takes the form
\ba
L&= & J'  [\sum_{\alpha=a,b} {\frac {1}{2 J' U}}\dot{\theta_\alpha}^2 +  [ \frac{J}{2J'(N-1)} (\theta_\alpha-\Phi_\alpha)^2 - \cos(\theta_\alpha)  ] \nonumber  \\
&+&\frac{J''}{J'}\frac{(\theta_a-\theta_b)^2}{2}] \; .
\ea
By applying the same procedure as in the previous section, we obtain the following Hamiltonian in the eigen-basis of the two-level systems of rings $a$ and $b$
\ba
H&=&H_a+H_b+\frac{J''}{J'}\sigma_x^{1}\sigma_x^{2}\langle \theta \rangle_{01}^2  \;  ,\\
H_\alpha&=&\epsilon\sigma_z^{\alpha}+(\frac{\Phi-\pi}{\delta}+\frac{J''\pi}{J'})\langle \theta \rangle_{01}\sigma_x^{\alpha} \; .
\ea
From this equations it follows that qubit-qubit interactions can be realized using our set-up. If we choose the tuning $\varepsilon\rightarrow0$ and $\Phi \rightarrow \pi-\frac{\delta J''\pi}{J'}$ the natural representation of a $(SWAP)^{\alpha}$ gate\cite{SWAP} can be obtained:
\be
U(\tau)=exp[-i\frac{J''}{J'}\sigma_x^{1}\sigma_x^{2}\tau],
\ee
where $\alpha=\frac{\tau J''}{ J'}$. A CNOT gate can be realized by using two $\sqrt{SWAP}$ gates. It is well known that one qubit rotations and a CNOT gate are sufficient to implement a set of universal quantum gates\cite{loss}.
}

{\bf{Acknowledgements}}\\
The authors are grateful to A. J. Leggett for his constant support since the early stages of this work and for a critical reading of the manuscript.
We thank F. Cataliotti, R. Fazio, F. Hekking, F. Illumninati, and G. Rastelli for discussions, and B. DeMarco for providing the original MRAF algorithm.
The work was financially supported by the National Research Foundation \& Ministry of Education, Singapore.
{D. Aghamalyan and Kwek L.-C. acknowledge financial support from Merlion Lumaton grant 2.08.11.}

{\bf Author contributions}\\
L.A. proposed the idea and developed it further together with D.A. and L.-C.K. H.C. proposed and carried out the experimental implementation with assistance from F.A. and advice from R.D. All authors discussed the results. L.A. and H.C. drafted the manuscript.\\

{\bf Additional information}\\

{\bf Competing financial Interests:} The authors declare no competing financial interests.



\newpage

\begin{appendix}
\begin{center}
{\bf SUPPLEMENTAL MATERIAL}
\end{center}

In the Appendix  \ref{sec:qubit_dynamics}, the derivation of the effective two-level dynamics of the system (single ring with a dimple) is provided.  In Appendix  
\ref{sec:real_dynamics}, we detail  on the analysis of the  dynamics of phase and population imbalances of  coupled persistent currents flowing in the  system, respectively. In the Appendix \ref{timeoflight}, details about time-of-flight density distributions plotted in Fig.4 are presented.
\maketitle

\section{Effective qubit dynamics}
\label{sec:qubit_dynamics}
In this section, we demonstrate how  the effective phase dynamics indeed defines a qubit.
To this end, we elaborate on the imaginary-time path integral of the partition function of the model Eq.(\ref{model}) in the limit of large fluctuations of the number of bosons at each site.  We first perform a local gauge transformation
$a_l\rightarrow  a_l e^{i l \Phi}$ eliminating the contribution of the magnetic field everywhere except at the weak link site where the phase slip is concentrated\cite{twisted boundary}). In the regime under scrutiny, the dynamics is governed by the Quantum-Phase Hamiltonian\cite{Fazio}
\begin{eqnarray}
H_{QP}=\sum_{i=0}^{N-2} \left [ U n_i^2 -J \cos\left ( \phi_{i+1,}-\phi_{i,}\right )\right ]   + \\
\left [ U n_{N-1}^2-J' \cos\left ( \phi_{0,}-\phi_{N-1,}-\Phi \right )\right ]
\end{eqnarray}
where $n_i$ and $\phi_i$ are conjugated variables and with $J=t \langle n \rangle $ and $J'=t' \langle n \rangle $.

The partition function of the model Eq.(\ref{model}) is
\begin{equation}
Z=Tr \left ( e^{-\beta H_{BH}} \right )  \propto \int D[\{ \phi_i \}]  e^{-S[\{ \phi_i \}]}
\end{equation}
where the effective action is
\begin{widetext}
\begin{eqnarray}
S[\{\phi_i\}]=\int d\tau \sum_{i=0 }^{N-2}  \left [ {\frac{1}{U}} (\dot{\phi}_{i})^2-J \cos\left ( \phi_{i+1,}-\phi_{i}\right )\right ]   + \left [ {\frac{1}{U}} (\dot{\phi}_{N-1})^2-J' \cos\left ( \phi_{0}-\phi_{N-1}-\Phi\right )\right ]
\end{eqnarray}
\end{widetext}

Because of the gauge transformations, the phase slip is produced only at the boundary. We define $\theta\doteq \phi_{N-1,}-\phi_{0,}$. The goal, now,  is to integrate out the phase variables in the bulk. To achieve the task, we observe that in the phase-slips-free-sites the phase differences are small, so the harmonic approximation can be applied:
\begin{eqnarray}
\sum_{i=0}^{N-2} \cos\left ( \phi_{i+1}-\phi_{i}\right ) \simeq \sum_{i=0}^{N-2}   {{\left ( \phi_{i+1}-\phi_{i}\right )^2}\over{2}}  \;.
\end{eqnarray}
In order to facilitate the integration in the bulk phases, we express the single $\phi_{0}$ and $\phi_{N-1}$ as: $\phi_{0}=\tilde{\phi}_{0}+\theta/2$, $\phi_{N-1}=\tilde{\phi}_{0}-\theta/2$. We observe that the sum of the quadratic terms above involves $N-1$ fields with periodic boundary conditions: $\{\tilde{\phi}_{0,}, \phi_{1},\dots,  \phi_{N-2} \} \equiv \{ \psi_{0},\psi_{1},\dots, \psi_{N-2}\}$, $\psi_{N-1}=\psi_{0}$. Therefore
\begin{eqnarray}
\sum_{i=0}^{N-2}  \left ( \phi_{i+1}-\phi_{i}\right )^2=\sum_{i=0}^{N-2}  \left ( \psi_{i+1}-\psi_{i}\right )^2 + \nonumber \\
{{1}\over{2}}\theta^2 +\theta \left (\psi_{N-2}-\psi_{1} \right ) \;.
\end{eqnarray}

The effective action, $S[\{\phi_i\}]$, can be split into two terms
$
S[\{\phi_i\}]=S_{1}[\theta] +S_{2}[\{\psi_{i}\}]
$
with
\begin{widetext}
\begin{eqnarray}
&&S_{1}[\theta]=\int d\tau \left [ {\frac{1}{U}} (\dot{\theta})^2+ {{J}\over{2}} \theta^2-J' \cos\left (\theta -\Phi \right )\right ]  \label{S_theta} \\
&& S_{2}[\{\psi_{i}\}, \theta]=\int d\tau \left \{{\frac{1}{U}} (\dot{\psi}_{0})^2+  \sum_{i=0}^{N-2} \left [ {\frac{1}{U}} (\dot{\psi}_{i})^2 +{{J}\over{2}}   \left ( \psi_{i+1}-\psi_{i}\right )^2 \right ]+ J  \theta \left (\psi_{N-2}-\psi_{1} \right )  \right \}
\label{S_coupling}
\end{eqnarray}
\end{widetext}
The integration of the fields $\psi_{i}$ proceeds according to the standard methods (see \cite{Hekking}). The fields that need to be integrated out are expanded in Fourier series ($N$ is assumed to be even):
$\psi_{l}=\psi_{0}+(-)^l \psi_{N/2}+\sum_{k=1}^{(N-2)/2} \left (\psi_{k} e^{{{2\pi i k l}\over{N-1}}}+c.c. \right )$,  with $\psi_{k}=a_{k}+ib_{k}$. The coupling term in Eq. (\ref{S_coupling}) involves only the imaginary part of $\psi_{k}$: $ \psi_{N-2}-\psi_{1}=\sum_k b_{k} \zeta_k$, being  $\displaystyle{\zeta_k={{4}\over{\sqrt{N-1}}} \sin \left ( {{2 \pi k}\over{N-1}}\right ) } $. Therefore:
\begin{widetext}
\begin{equation}
S_{2}[\{\psi_{i}\}, \theta] =\int d\tau {{1}\over{U}}\sum_k (\dot{a}_{k})^2 +\omega_k^2 a_{k}^2 +
\int d\tau {{1}\over{U}}\sum_k (\dot{b}_{k})^2 +\omega_k^2 b_{k}^2+ J U \zeta_k\theta   b_{k}
\end{equation}
\end{widetext}
where $ \omega_k= \sqrt{2 J U \left [ 1-\cos{ \left ( {{2\pi k}\over{N-1}} \right )} \right ] } $. The integral in $\{a_{k} \} $ leads to a Gaussian path integral; it does not contain the interaction with $\theta$, and therefore brings a prefactor multiplying the effective action, that does not affect the dynamics. The integral in $\{b_{k} \} $ involves  the interaction and therefore leads to a non local kernel in the imaginary time: $\int d\tau d\tau' \theta(\tau) G(\tau-\tau')\theta(\tau')$. The explicit form of $G(\tau-\tau') $ is obtained by expanding  $\{b_{k} \} $ and $\theta$ in Matsubara frequencies $\omega_l$. The corresponding Gaussian integral yields  to the
\begin{equation}
\int D[b_{k}] e^{-\int d \tau S_{02}}\propto \exp{\left (-\beta U J^2 \sum_{l=0}^{\infty} \tilde{Y}(\omega_l) |\theta_l |^2  \right )}
\end{equation}
with $\tilde{Y}(\omega_l)=\sum_{k=1}^{(N-2)/2} {{\zeta_k^2}\over{\omega_k^2+\omega_l^2}}$. The $\tau=\tau'$ term is extracted by summing and subtracting $\tilde{Y}(\omega_l=0)$; this compensates the second term in Eq.(\ref{S_theta}).

The effective action finally reads as
\begin{widetext}
\begin{equation}
S_{eff}=\int_0^\beta d \tau \left [ {\frac {1}{2 U}}  \dot{\theta}^2 + U(\theta) \right ]
- \frac{J}{2 U(N-1)} \sum   \int d \tau d \tau' \theta(\tau)G_(\tau-\tau') \theta(\tau')
\label{effective_single}
\end{equation}
\end{widetext}
where
\begin{eqnarray}
\displaystyle{U(\theta)\doteq \frac{J}{N-1} (\theta-\Phi)^2 -J' \cos \theta}
\end{eqnarray}
plotted in Fig.\ref{effective_single_potential}.
The kernel in the non-local term  is given by
\begin{equation}
G(\tau)=\sum_{l=0}^\infty  \sum_{k=1}^{{{N-2}\over{2}}} {\frac{\omega_l^2 \left (1+\cos[{{2 \pi k}\over{N-1}}]\right )}{2 JU(1-\cos[{{2 \pi k}\over{N-1}}])+\omega_l^2}} e^{i\omega_l \tau}\; .
\end{equation}
\begin{figure}
{\includegraphics*[scale = 0.5]{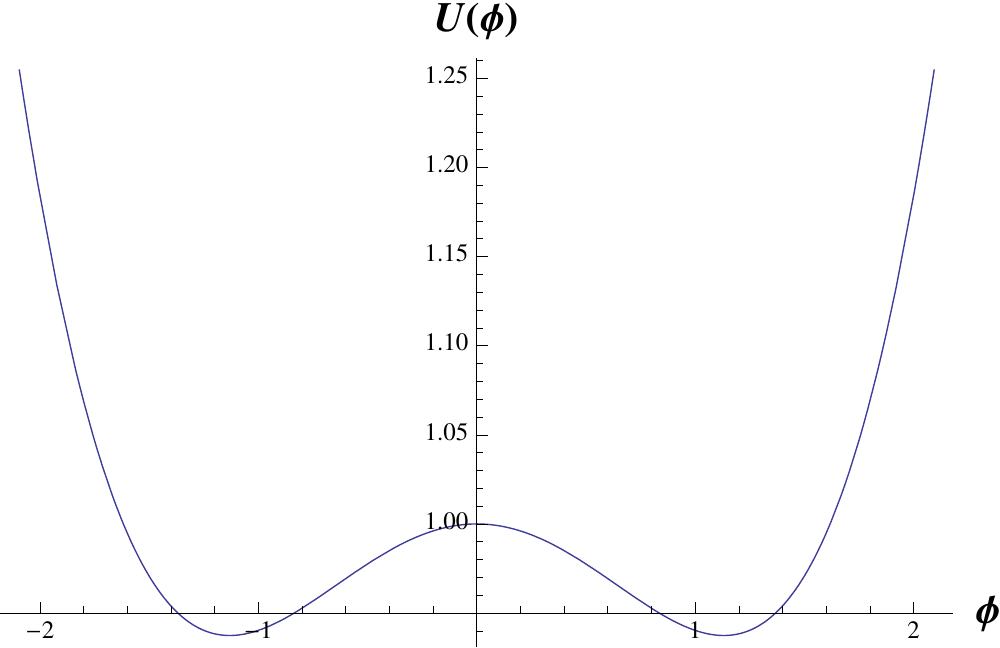}}
\caption{ The double well potential providing the  single-ring-qubit for  $J/[J'(N-1)]=0.4$ and $ \Phi=\pi$}
\label{effective_single_potential}
\end{figure}
\begin{figure}
{\includegraphics*[scale = 0.5]{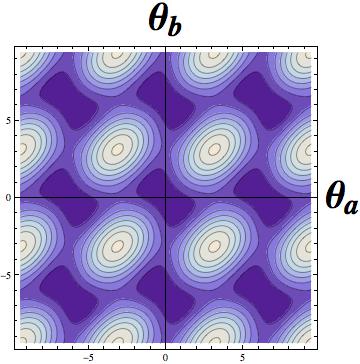}}
{\includegraphics*[scale = 0.5]{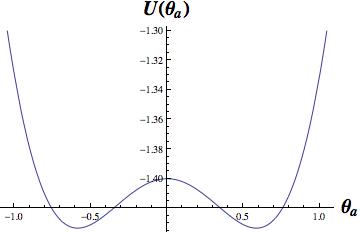}}
\caption{ (Left) The effective potential landscape providing  the two-rings-qubit.
(Right) The double well for $\theta_a=-\theta_b$.
The parameters are $\tilde{J}/J=0.8$ and $\Phi_a-\Phi_b=\pi$.}
\label{effective_tworings_potential}
\end{figure}

The external bath vanishes in the thermodynamic limit and the effective action reduces to the Caldeira-Leggett one \cite{Hekking}.
Finally it is worth noting that the case of a single junction needs a specific approach but it can be demonstrated consistent with Eq.(\ref{effective_single}).

For the two rings with tunnel coupling, a similar procedure is applied. The effective action (4) is obtained under the assumption
that the two rings are weakly coupled and that $U/J<<1$. The effective potential  (Eq.(5) of the main manuscript) for the two-rings-qubit is displayed in Fig.\ref{effective_tworings_potential}\cite{davit}.

\section{Real time dynamics: Two  coupled Gross-Pitaevskii equations}
\label{sec:real_dynamics}

In this section we study the dynamics of the  number and phase imbalance  of  two bose-condensates  confined  in the ring shaped potential (see also \cite{davit}). A single-species bosonic condensate  is envisaged to be loaded in the setup described above.  Our system is thus governed by a Bose-Hubbard ladder Hamiltonian
\begin{equation}
 H_{BH}=H_{a}+H_{b}+H_{int} - \sum_{=a,b}\sum_{i=0}^{N-1}\mu_{}\hat{n}_{i}^{}
 \label{model}
 \end{equation}
 with
\begin{eqnarray}
&&H_{a}=-t\sum_{i=0}^{N-1}(e^{i\Phi_{a}/N}a_{i}^{\dag}a_{i+1}+h.c.)+\frac{U}{2}\sum_{i=1}^{N}\hat{n}_{i}^{a}(\hat{n}_{i}^{a}-1) \nonumber  \\
%
&&H_{b}=-t\sum_{i=0}^{N-1}(e^{i\Phi_{b}/N}b_{i}^{+}b_{i+1}+h.c.)+\frac{U}{2}\sum_{i=1}^{N}\hat{n}_{i}^{b}(\hat{n}_{i}^{b}-1) \nonumber \\
 %
 &&H_{int}=-g\sum_{i=0}^{N-1}(a_{i}^{\dag}b_{i}+b_{i}^{\dag}a_{i})
 \label{model_int}
 \end{eqnarray}
where $H_{a,b}$ are the Hamiltonians of the condensates in the rings $a$ and $b$ and the $H_{int}$ describes the interaction between rings. Operators $\hat{n}_{i}^{a}=a_{i}^{\dag}a_{i},
\hat{n}_{i}^{b}= b_{i}^{\dag}b_{i}$ are the particle number operators for the lattice site $i$. Operators $a_{i}$ and $b_{i}$ obey the standard bosonic commutation relations. The parameter $t$ is the tunneling rate within lattice neighboring sites, and $g$ is the tunneling rate between the rings. The on-site repulsion between two atoms  is quantified by  $U=\frac{4\pi a_{s} \hbar^{2}}{m}\int |\textit{w}(\textbf{x})|^4 d^3\textbf{x}$, where $a_{s}$ is the $s$-wave scattering length of the atom and $|\textit{w}(\textbf{x})|$ is a single-particle Wannier function.
Finally, the phases $\Phi_a$ and $\Phi_b$ are the phase twists responsible for the currents flowing along the rings.
%
They can be expressed through vector potential of the so-called synthetic gauge fields in the following way: $\Phi_{a}/N=\int_{x_{i}}^{x_{i+1}}\emph{\textbf{A(z)}}d\emph{\textbf{z}}$, $\Phi_{b}/N=\int_{x_{i}}^{x_{i+1}}\emph{\textbf{B(z)}}d\emph{\textbf{z}}$, where $\emph{\textbf{A(z)}}$ and $\emph{\textbf{B(z)}}$ are generated vector potentials in the rings $a$ and $b$, respectively.  We would like to emphasize, that the inter-ring hopping element $g$ is not affected by the Peierls substitution because  the synthetic gauge field is assumed to  have components longitudinal  to the rings only.

To obtain the Gross-Pitaevskyi, we assume that the system is described by a Bose-Hubbard ladder Eqs.(\ref{model}), is in a superfluid  regime, with   negligible quantum fluctuations.
The order parameters can be defined as the expectation values of bosonic operators in the Heisenberg picture:
\begin{eqnarray}
 \varphi_{a,i}(s)=\langle a_{i}(s)\rangle,\varphi_{b,i}(s)=\langle b_{i}(s)\rangle \;,
 \label{app0}
\end{eqnarray}
implying that the  Heisenberg equations for  the operators $a_{i}$ and $b_{i}$ are simplified into the  Gross-Pitaevskii equations for the corresponding expectation values:
\begin{eqnarray}\nonumber
i\hbar\frac{\partial \varphi_{a,i} }{\partial s}=-t(e^{i\Phi_{a}/N}\varphi_{a,i+1}+e^{-i\Phi_{a}/N}\varphi_{a,i-1}) \\
+U|\varphi_{a,i}|^{2}\varphi_{a,i}-\mu_{a}\varphi_{a,i}-g\varphi_{b,i}
\label{eq1}
\end{eqnarray}
\begin{eqnarray}\nonumber
i\hbar\frac{\partial \varphi_{b,i} }{\partial s}=-t(e^{i\Phi_{b}/N}\varphi_{b,i+1}+e^{-i\Phi_{b}/N}\varphi_{b,i-1}) \\
+U|\varphi_{b,i}|^{2}\varphi_{b,i}-\mu_{b}\varphi_{b,i}-g\varphi_{a,i}
\label{eq2}
\end{eqnarray}
We assume that
$\varphi_{a,i+1}-\varphi_{a,i}=\frac{\varphi_{a}(s)}{\sqrt{N}}$  and  $\varphi_{b,i+1}-\varphi_{b,i}=\frac{\varphi_{b}(s)}{\sqrt{N}}$
for all $i,j=0,..,N$,
where $N$ is a total number of ring-lattice sites.
From Eqs.(\ref{eq1}) and (\ref{eq2}) we obtain
\begin{eqnarray}\nonumber
i\hbar\frac{\partial \varphi_{a} }{\partial s}=-2t\cos{(\Phi_{a}/N)}\varphi_{a}
+\frac{U}{N}|\varphi_{a}|^{2}\varphi_{a} \\
-\mu_{a}\varphi_{a}-g\varphi_{b}
\label{eq3}
\end{eqnarray}
\begin{eqnarray}\nonumber
i\hbar\frac{\partial \varphi_{b} }{\partial s}=-2t\cos{(\Phi_{b}/N)}\varphi_{b}
+\frac{U}{N}|\varphi_{b}|^{2}\varphi_{b} \\
-\mu_{b}\varphi_{b}-g\varphi_{a}
\label{eq4}
\end{eqnarray}
Employing the standard phase-number representation:  $\varphi_{a,b}=\sqrt{N_{a,b}} e^{i\theta{a,b}}$,  two pairs of equations are obtained for imaginary and real parts:
\begin{eqnarray}\nonumber
\hbar\frac{\partial N_{a} }{\partial s}=-2g\sqrt{N_{a}N_{b}}\sin{(\theta{b}-\theta{a})}
\end{eqnarray}
\begin{eqnarray}
\hbar\frac{\partial N_{b} }{\partial s}=2g\sqrt{N_{a}N_{b}}\sin{(\theta{b}-\theta{a})}
\label{eq5}
\end{eqnarray}
\begin{eqnarray}\label{eq6}
\hbar\frac{\partial \theta{a} }{\partial s}=-2t\cos{\Phi_{a}/N}-\frac{U N_{a}}{N}+\mu_{a}+g\sqrt\frac{N_{b}}{N_{a}}\cos{(\theta{b}-\theta{a})} \nonumber \\
\hbar\frac{\partial \theta{b} }{\partial s}=-2t\cos{\Phi_{b}/N}-\frac{U N_{b}}{N}+\mu_{b}+g\sqrt\frac{N_{a}}{N_{b}}\cos{(\theta{b}-\theta{a})} \nonumber \\
\end{eqnarray}
From Eqs.(\ref{eq5}) it results that $\frac{\partial N_{a} }{\partial s}+\frac{\partial N_{b} }{\partial s}=0$, reflecting the conservation of the total bosonic number $N_T=N_a+N_b$.
From equations (\ref{eq5}) and (\ref{eq6}) we get
\begin{equation}
\frac{\partial z }{\partial \tilde {s}}=-\sqrt{1-z^{2}}\sin{\Theta}
\label{dynamical_equation1}
\end{equation}
\begin{equation}
\frac{\partial \Theta }{\partial \tilde {s}
}=\Delta+\lambda \rho z+\frac{z}{\sqrt {1-z^{2}}}\cos{\Theta}
\label{dynamical_equation2}
\end{equation}
where we introduced  new variables:the dimensionless time $ 2gs/\hbar\rightarrow \tilde{s}$,the population imbalance $z(\tilde{s})=(N_{b}-N_{a})/(N_{a}+N_{b})$ and the phase difference between the two condensates $\Theta(\tilde{s})=\theta{a}-\theta{b}$.
It is convenient to characterize  the system with a new set of parameters: external driving force $\Delta=(2t(\cos{\Phi_{a}/N}-\cos{\Phi_{b}/N})+\mu_{b}-\mu_{a})/2g$,  effective scattering wavelength $\lambda =U/2g$ and total bosonic density $\rho=N_{T}/N$.
The exact  solutions of  Eqs.(\ref{dynamical_equation1}) and (\ref{dynamical_equation2})  in terms of elliptic functions\cite{Smerzi}  can be adapted to our case\cite{davit}.
The equations can be derived as Hamilton equations with
\begin{equation}
H(z(\tilde {s}),\Theta(\tilde {s}))=\frac{\lambda \rho z^2}{2}+\Delta z-\sqrt{1-z^2}cos\Theta,
\label{eff ham}
\end{equation}
by considering $z$ and $\phi$ as conjugate variables.
Since  the energy of the   system is conserved,   $H(z(\tilde {s}),\Theta(\tilde {s}))=H(z(0),\Theta(0))=H_{0}$.
Combining Eqs.(\ref{dynamical_equation1}) and  (\ref{eff ham}),  $\Theta$ can be eliminated, obtaining
\begin{equation}
\dot{z}^2+[\frac{\lambda \rho z^2}{2}+\Delta z-H_0]^2=1-z^2,
\label{z equation}
\end{equation}
that is solved  by  quadratures:
\begin{equation}
\frac{\lambda\varrho \tilde {s}}{2}=\int_{z(0)}^{z(\tilde {s})}\frac{dz}{\sqrt{f(z)}} \; ,
\label{z solution}
\end{equation}
where $f(z)$ is the  following quartic equation
\begin{equation}
f(z)=\big(\frac{2}{\lambda \rho}\big)^2(1-z^2)-\big[z^2+\frac{2z\Delta}{\lambda \rho}-\frac{2H_0}{\lambda \rho}\big]^2 \; .
\label{quartic}
\end{equation}

There are two different cases: $\Delta=0$ and $\Delta\neq 0$.

{ {I)}}  $\Delta=0.$ --
In this case the solution for the $z(t)$ can be expressed in terms of  'cn' and 'dn' Jacobian elliptic functions as(\cite{Smerzi}):
\begin{eqnarray}
z(\tilde {s})&=&C cn[(C\lambda \rho/k(\tilde {s}-\tilde {s}_{0}),k)] \ \ for \ \ 0<k<1 \nonumber \\
&=& C sech(C\lambda \rho(\tilde {s}-\tilde {s}_{0})), \ \ for \ \ k=1 \nonumber \\
&=&C dn[(C\lambda \rho/k(\tilde {s}-\tilde {s}_{0}),1/k)] \ \ for \ \ k>1; \label{sol1} \\
k&=&\big(\frac{C\lambda \rho}{\sqrt{2}\zeta(\lambda \rho)}\big)^2=\frac{1}{2}\big[1+\frac{(H_{0}\lambda \rho-1)}{(\lambda \rho)^2+1-2H_{0}\lambda \rho}\big],\label{el mod}
\end{eqnarray}
where
\begin{eqnarray}
C^2=\frac{2}{(\lambda \rho)^2}((H_{0}\lambda \rho-1)+\zeta^2),\nonumber \\
^2=\frac{2}{(\lambda \rho)^2}(\zeta^2-(H_{0}\lambda \rho-1)), \nonumber \\
\zeta^2(\lambda \rho)=2\sqrt{(\lambda \rho)^2+1-2H_{0}\lambda \rho},\label{paramet}
\end{eqnarray}
and $\tilde {s}_{0}$  fixing $z(0)$. Jacobi functions are defined in terms of the incomplete elliptic integral of the first kind  $F(\phi,k)=\int_{0}^{\phi}d\theta(1-k \sin^2{\theta})^{-1/2}$ by the following expressions: $sn(u|k)=\sin{\phi},cn(u|k)=\cos{\phi}$ and $dn(u|k)=(1-k \sin^2{\phi})^{1/2}$~\cite{Abramowitz}. The Jacobian elliptic functions $sn(u|k)$, $cn(u|k)$ and $dn(u|k)$ are periodic in the argument $u$ with period $4K(k)$, $4K(k)$ and $2K(k)$,  respectively, where  $ K(k)=F(\pi/2,k)$ is the complete elliptic integral of the first kind. For small elliptic modulus $k\simeq 0$,  such functions  behave as trigonometric functions; for  $k\simeq 1$, they behave as hyperbolic functions. Accordingly, the  character of the solution of Eqs.(\ref{dynamical_equation1}) and (\ref{dynamical_equation2}) can be   oscillatory or exponential,  depending  on $k$.
For  $k\ll 1$, $cn(u|k)\approx\cos{u}+0.25 k (u-\sin{(2u)}/2)\sin{u}$ is almost sinusoidal and the population imbalance is oscillating around a zero average value. When $k$ increases, the  oscillations  become non-sinusoidal and for $1-k\ll 1$ the time evolution  is non-periodic: $cn(u|k)\approx\sec{u}-0.25 (1-k) (\sinh{(2u)}/2-u)\tanh{u}\sec{u}$. From the last expression,  we can see that at $k=1$, $cn(u|k)=\sec{u}$ so oscillations are exponentially suppressed and $z(\tilde{s})$ taking $0$ asymptotic value. For the values of the $k>1$ such that $[1-1/k]\ll 1$ $z(s)$ is still non-periodic and is given by: $dn(u|1/k)\approx\sec{u}+0.25 (1-1/k) (\sinh{(2u)}/2+u)\tanh{u} \sec{u}$.  Finally when $k\gg 1$ than the behavior switches to   sinusoidal again,  but $z(\tilde{s})$ does oscillates around  a non-zero average: $dn(u|1/k)\approx 1-\sin^2{u}/2k$. This phenomenon accounts for the MQST.
%

%
%

{ {II)}}$\Delta \neq 0 $.--
In this case $z(s)$ is expressed in terms of the Weierstrass elliptic function(\cite{Kenkre,Smerzi})
\begin{equation}
z(\tilde {s})=z_1+\frac{f^{\prime}(z_1)/4}{\varrho(\frac{\lambda \rho}{2}(\tilde {s}-\tilde {s}_{0});g_{2},g_3)-\frac{f^{\prime\prime}(z_1)}{24}} \; ,
\label{weier sol}
\end{equation}
where $ f(z) $ is given by an expression (\ref{quartic}), $z_1$ is a root of quartic $f(z)$ and $\tilde {s}_{0}=(2/\lambda \rho)\int_{z_1}^{z(0)}\frac{dz^{\prime}}{\sqrt{f(z^{\prime})}}$. For $\sin{\Theta_0}=0$ (which is the case discussed in the text),  $z_1=z_0$ and consequently $s_0=0$.
The Weierstrass elliptic function can be given as the inverse of an elliptic integral $\varrho(u;g_2,g_3)=y$, where
\begin{equation}
u=\int_y^\infty \frac{ds}{\sqrt{4s^3-g_2 s-g_3}} \; .
\end{equation}
The constants $g_2$ and $g_3$ are  the characteristic  invariants  of $\varrho$:
\begin{eqnarray}\nonumber
g_2 &=& -a_4-4a_1a_3+3a_2^2 \\
g_3 &=&-a_2 a_4+2a_1 a_2  a_3-a_2^3+a_3^2-a_1^2a_4,
\end{eqnarray}
where the coefficients $a_i$, where $i=1,..,4$, are given as
\begin{eqnarray}\nonumber
a_1&=&-\frac{\Delta}{\lambda \rho};a_2=\frac{2}{3(\lambda \rho)^2}(\lambda \rho H_0-(\Delta^2+1)) \\
a_3&=&\frac{2H_0\Delta}{(\lambda \rho)^2};a_4=\frac{4(1-H_0^2)}{(\lambda \rho)^2}
\end{eqnarray}
In the present case ($\Delta \neq 0 $),  the discriminant
\begin{equation}
\delta=g_2^3-27g_3^2
\label{discrim}
\end{equation}
 of the cubic $h(y)=4y^3-g_2y-g_3$ governs the behavior of the Weierstrass elliptic functions (we contrast with
 the case  $\Delta=0$, where  the dynamics is governed by the elliptic modulus $k$).
%
If   $g_2<0,g_3>0 $ then(\cite{Abramowitz})
\begin{equation}
z(\tilde {s})=z_1+\frac{f^{\prime}(z_1)/4}{c+3c\sinh^{-2}{[\frac{\sqrt{3c}\lambda \rho}{2}(\tilde {s}-\tilde{s}_{0})]}-\frac{f^{\prime\prime}(z_1)}{24}}
\label{sol-delta=0(1)} \; .
\end{equation}
Namely, the  oscillations of $z$ are exponentially suppressed  and the population imbalance decay (if $z_0>0$) or saturate (if $z_0<0$) to the asymptotic value given by
$z(\tilde{s})=z_1+\frac{f^{\prime}(z_1)/4}{c-f^{\prime\prime}
 (z_1)/24}$. \\
If  $ g_2>0,g_3>0$ then(\cite{Abramowitz})
\begin{equation}
z(\tilde {s})=z_1+\frac{f^{\prime}(z_1)/4}{-c+3c\sin^{-2}{[\frac{\sqrt{3c}\lambda \rho}{2}(\tilde {s}-\tilde {s}_{0})]}-\frac{f^{\prime\prime}(z_1)}{24}}
\label{sol-delta=0(2)}  \; ,
\end{equation}
where $c=\sqrt{g_2/12}$.
We see that the population imbalance  oscillates around a non-zero average value $\overline{z}\doteq z_1+\frac{f^{\prime}(z_1)/4}{2(2c-f^{\prime\prime}(z_1)/24)}$,  with  frequency
$\omega=2g \sqrt{3c}\lambda\rho$.

We  express the Weierstrass function in terms of Jacobian elliptic functions. This leads to  significant  simplification  for the analysis of these regimes.

For $\delta>0$, it results
\begin{equation}
z(\tilde {s})=z_1+\frac{f^{\prime}(z_1)/4}{e_3+\frac{e_1-e_3}{sn^{2}[\frac{\lambda \rho \sqrt{e_1-e_3}}{2}(\tilde {s}-\tilde {s}_{0}),k_1]}-\frac{f^{\prime\prime}(z_1)}{24}}
\label{sol-delta>0}  \; ,
\end{equation}
where $k_1=\frac{e_{2}-e_{3}}{e_{1}-e_{3}} $ and $e_{i}$ are  solutions of the cubic equation $h(y)=0$.
 In this case   the population imbalance oscillates  about the average value $\overline{z}=z_1+\frac{f^{\prime}(z_1)/4}{2(e_1-f^{\prime\prime}(z_1)/24)}$.

The asymptotics of the solution is extracted through:  $k\ll 1$, $sn(u|k)\approx\sin{u}-0.25 k (u-\sin{(2u)}/2)\cos{u}$. When $k$ increases oscillations starting to become non-sinusoidal and when $1-k\ll 1$ it becomes non-periodic and takes form: $cn(u|k)\approx\tanh{u}-0.25 (1-k) (\sinh{(2u)}/2-u)\sec^2{u}$.

For $\delta<0$ the following expression for $z(s)$ is obtained:
\begin{equation}
z(\tilde {s})=z_1+\frac{f^{\prime}(z_1)/4}{e_2+H_2\frac{1+cn[\lambda \rho \sqrt{H_2}(\tilde {s}-\tilde {s}_{0}),k_2]}{1-cn[\lambda \rho \sqrt{H_2}(\tilde {s}-\tilde {s}_{0}),k_2]}-\frac{f^{\prime\prime}(z_1)}{24}}  \; ,
\label{sol-delta<0}
\end{equation}
where $k_2=1/2-\frac{3e_{2}}{4H_{2}}$ and $H_{2}=\sqrt{3e_{2}^2-\frac{g_{2}}{4}}$.The asymptotical behavior of the function $cn(u|k)$ has been discussed in the previous subsection.
As it it seen from this expression $z(\tilde {s})$ oscillates  about the average value ${\overline{z}}=z_1+\frac{f^{\prime}(z_1)/4}{2(e_2-f^{\prime\prime}(z_1)/24)}$.


\subsection{\protect\normalsize Population imbalance and oscillation frequencies in the limit $\lambda \rho\ll 1$ }

{I-B} $\Delta=0.$--
The qualitative behavior of the dynamics for this sub-case depends  on the elliptic modulus  $k$ which is given by Eq.(\ref{el mod}).
 For $\lambda \rho\ll 1$
 \begin{equation}
k=z(0)\lambda \rho(1-\frac{\lambda \rho}{2}\sqrt{1-z(0)^2})
\end{equation}
implying that $ k\approx 0$; therefore $z(t)$ displays only one regime given by
 \begin{eqnarray}
z(\tilde {s}) &
\simeq & z(0)(\cos{\omega (\tilde {s}-\tilde {s}_{0})} \\ \nonumber
&+&   \frac{k}{4}(\omega (\tilde {s}-\tilde {s}_{0})-\sin{2 \omega (\tilde {s}-\tilde {s}_{0})})
\sin{ \omega (\tilde {s}-\tilde {s}_{0})}) \; .
\end{eqnarray}
 where $\omega\simeq 2g(1+\frac{\lambda}{2} \rho\sqrt{1-z(0)^2})$ and $\tilde {s}_{0}$ is  fixing initial condition.
Therefore, in this regime the population imbalance is characterized by  almost sinusoidal oscillations about zero average-- see the inset of Fig. 3 of the main part of the material. 

{ II-B} $\Delta\neq 0.$--
In this case, the  behavior of $z(t)$ is governed by the discriminant $\delta$ of the  cubic equation Eq.(\ref{discrim}).
There are two different regimes depending on the initial value of the population imbalance which are given by the value of $\delta$.
All the regimes can be discussed by expressing the Weierstrass function in Eq.(\ref{weier sol}) using Jacobian elliptic functions.
In the limit of $\delta=0$, the population imbalance is
\begin{equation}
z(\tilde {s})=z(0)+\frac{f'[z(0)]/4}{-c+3c[\sin
{(-\sqrt{3c}\frac{\lambda \rho}{2} \tilde {s}})]^{-2}-f''[z(0)]/24}  \; .
\end{equation}
For the parameters discussed in Fig. 3 of the main article, $ f'[z(0)]\sim 10^{-14}$; therefore the population imbalance is  constant due to the same reason discussed for  $D=0$ above.
In the limit of $\delta<0$, the population imbalance is
\begin{equation}
z(\tilde {s})=z(0)+\frac{f'[z(0)]/4}{e_{2}+H_{2}\frac{1+\cos{(\lambda \rho  \sqrt{H_{2}\tilde {s}}})}{1-\cos{(\lambda \rho  \sqrt{H_{2}\tilde {s}}})}-f''[z(0)]/24}  \; ,
\label{saturation}
\end{equation}
where $e2,H_{2}$ are defined in the Appendix A. Eq.(\ref{saturation}) is correct when $1/2-3e_{2}/4H_{2}\simeq 0$( for the parameters considered in the article $m\simeq 10^{-7}$).
As one sees from this formula, the population imbalance displays an oscillating behavior  around a  non-zero average (MQST regime) with frequency given by
\begin{equation}
\omega=2g(\sqrt{1+\Delta^2}+\frac{(z(0)\Delta-\sqrt{1-z(0)^2})(2\Delta^2-1)}{2(1+\Delta^2)^{3/2}}\lambda\rho) \; .
\end{equation}
 This two regimes  are shown  in Fig. 3 of the main article. 


\section{Time of flight}
\label{timeoflight}

In this section the density of momentum distribution  which can be observed in the time of flight type of measurement for a Bose-Hubbard ladder model Eq.(\ref{model}) is derived. The density of momentum distribution is given by

\be
\rho(\textbf{k})=\int d^3x \int d^3x' \langle \Psi(\textbf{x})^{\dag} \Psi(\textbf{x}')\rangle e^{i \textbf{k}(\textbf{x}-\textbf{x}') }  \; ,
\label{momentum dist}
\ee
where $\Psi^{\dag}(\textbf{x})$ and $\Psi(\textbf{x}')$ are bosonic field-operators.

Let us express them through Wannier functions:
\be
\Psi(\textbf{x})=\sum_{i=0}^{N-1}[w(\textbf{x}-\textbf{r}_i)e^{i\varphi_i^{a}}a_i+w(\textbf{x}-\textbf{r}_i)e^{i\varphi_i^{b}}b_i]  \; ,
\label{wannier}
\ee
where  exponential factors arise from the Peierls substitution  and they are given by $\varphi_{i+1}^{a}-\varphi_{i}^{a}=2\pi\Phi_a/L^2$ and $\varphi_{i+1}^{b}-\varphi_{i}^{b}=2\pi\Phi_b/L^2$, where $\Phi_a$ and $\Phi_b$ are the fluxes induced in the rings $a$ and $b$ respectively. After substituting Eq.(\ref{wannier}) into Eq.(\ref{momentum dist}) and making change of variables $\textbf{z}=\textbf{x}-\textbf{r}_i$,$\textbf{z}'=\textbf{x}'-\textbf{r}_i$,  we get
\ba
\nonumber
\rho(\textbf{k})=\sum_i\sum_j[|w(\textbf{k})|^2(e^{i(\varphi_j^{a}-\varphi_i^{a})}\langle a^{\dag}_ia_j \rangle+e^{i(\varphi_j^{b}-\varphi_i^{b})}\langle b^{\dag}_ib_j \rangle \\
+e^{i(\varphi_j^{a}-\varphi_i^{b})}\langle b^{\dag}_ia_j \rangle+e^{i(\varphi_j^{b}-\varphi_i^{a})}\langle a^{\dag}_ib_j \rangle)]e^{i \textbf{k}(\textbf{r}_i-\textbf{r}_j)}  \; .
\ea
We  note that $z_i-z_j=0$ for $i$ and $j$ belonging to the same ring;  otherwise $z_i-z_j=\pm D$, $D$ being the distance between the rings.Therefore, the   momentum distribution reads 
\ba\nonumber
\rho(\textbf{k})&=&|w(\textbf{k})|^2[\sum_{i\in a}\sum_{j\in a}(e^{i((\varphi_j^{a}-\varphi_i^{a})+\textbf{k}_{\parallel}\cdot \textbf{x}_{\parallel})}\langle a^{\dag}_ia_j \rangle\\ \nonumber
&+&\sum_{i\in b}\sum_{j\in b}e^{i((\varphi_j^{b}-\varphi_i^{b})+\textbf{k}_{\parallel}\cdot \textbf{x}_{\parallel})}\langle b^{\dag}_ib_j \rangle \\ \nonumber
&+&\sum_{i\in a}\sum_{j\in b}e^{i((\varphi_j^{b}-\varphi_i^{a})+\textbf{k}_{\parallel}\cdot \textbf{x}_{\parallel}+k_z D)}\langle b^{\dag}_ia_j \rangle\\
&+&\sum_{i\in b}\sum_{j\in a}e^{i((\varphi_j^{a}-\varphi_i^{b})\textbf{k}_{\parallel}\cdot \textbf{x}_{\parallel}-k_z D)}\langle a^{\dag}_ib_j \rangle)] \; ,
\ea
where $w(\textbf{k})$ are  Wannier functions in the momentum space (that we considered identical for the two rings),
$\textbf{k}_{\parallel}\cdot \textbf{x}_{\parallel}\doteq k_{x}(x_{i}-x_{j})+ k_{y}(y_{i}-y_{j}) $,
$x_{i}=\cos{\phi_i}$,  $y_{i}=\sin{\phi_i}$ fix the positions of the ring wells in the three dimensional space, $\phi_i=2\pi i /N$ being lattice sites  along the rings.
Then we transform annihilation and creation operators to the momentum space $a_i=1/\sqrt{N}\sum_q e^{i \phi_i q }a_q$ and $b_i=1/\sqrt{N}\sum_q e^{i \phi_i q }b_q$. We also take into account that $\varphi_i^{a}=2\pi i\Phi_a/N$ and $\varphi_i^{b}=2\pi i\Phi_b/N$ for $i=0,..,N-1$. Finally, we get  (Eq.7)

\ba \label{density}
&&\rho(\textbf{k})={ {|w(k_x, k_y, k_z)|^{2}}\over{N} }\sum_{i=0}^{N-1}\sum_{j=0}^{N-1}\sum_{q\in \{ 2\pi n/N\}} \\
&& \left [ \cos{ [\textbf{k}_{\parallel}\cdot \textbf{x}_{\parallel} +(q+\frac{\Phi_{a}}{N}) (\phi_i-\phi_j)] } \langle a_{q}^{\dag}a_{q} \rangle +   \nonumber  \right .\\
&& \left. \cos{ [\textbf{k}_{\parallel}\cdot \textbf{x}_{\parallel}+(q+\frac{\Phi_{b}}{N}) (\phi_i-\phi_j)] } \langle b_{q}^{\dag}b_{q}  \rangle +   \nonumber  \right .\\
 && \left .2\cos{ [\textbf{k}_{\parallel}\cdot \textbf{x}_{\parallel} +k_{z}D+(q+\frac{\Phi_{a}}{N}) \phi_i-(q+\frac{\Phi_{b}}{N})\phi_j) ]}
\langle a_{q}^{\dag}b_{q}  \rangle \right ] \; . \nonumber
\ea

\subsection{Expectation values for $U=0$}
In the following, we provide the details of the calculations of the expectation values entering the Eq.(\ref{density}), for  $U=0$. 

The Hamiltonian in the Fourier space reads
%
\ba 
H_{BH}=\sum_{k}[-2t\cos{\tilde{k}_a}a_{k}^{\dag}a_{k}-2t\cos{\tilde{k}_b}b_{k}^{+}b_{k}
-g(a_{k}^{\dag}b_{k}+b_{k}^{\dag}a_{k})]
\label{Ham in k rep}
\ea
We perform a Bogolubov rotation
\ba \nonumber
a_k=\sin{\theta_k}\alpha_k+\cos{\theta_k}\beta_k \\
b_k=\cos{\theta_k} \alpha_k -\sin{\theta_k}\beta_k
\label{bogoliubov transform}
\ea
The Hamiltonian Eq.(\ref{Ham in k rep}) can be  diagonalized choosing  $\tan{2\theta_k}=g/t(\cos{\tilde{k}_a}+\cos{\tilde{k}_b})$:
\ba
H_{BH}&=&\sum_{k}[\varepsilon_{\alpha}(k)\alpha_{k}^{\dag}\alpha_{k}+\varepsilon_{\beta}(k)\beta_{k}^{+}\beta_{k}]\\ \nonumber
\varepsilon_{\alpha,\beta}(k)&=&-t(\cos{\tilde{k}_a+\cos{\tilde{k}_b})} \\ \nonumber
&\mp&\sqrt{g^2+t^2(\cos{\tilde{k}_a}-\cos{\tilde{k}_b})^2}
\ea
 where $\tilde{k}_a=k+\Phi_{a}/N,\tilde{k}_b=k+\Phi_{b}/N$ and $\pm$
 corresponds to the $\alpha$ and $\beta$ respectively.

The correlation functions result
  \ba
\langle a_{k}^{\dag}a_{k}  \rangle&=&\sin^{2}{\theta_k}\langle \alpha_{k}^{\dag}\alpha_{k}\rangle +\cos^{2}{\theta_k}\langle \beta_{k}^{\dag}\beta_{k}\rangle\\ \nonumber
\langle b_{k}^{\dag}b_{k}  \rangle&=&\cos^{2}{\theta_k}\langle \alpha_{k}^{\dag}\alpha_{k}\rangle +\sin^{2}{\theta_k}\langle \beta_{k}^{\dag}\beta_{k}\rangle\\ \nonumber
\langle a_{k}^{\dag}b_{k}  \rangle=\langle b_{k}^{\dag}a_{k}  \rangle&=&\frac{\sin{2\theta_k}}{2}(\langle \alpha_{k}^{\dag}\alpha_{k}\rangle -\langle \beta_{k}^{\dag}\beta_{k}\rangle)\\ \nonumber
\ea
where  $\langle \alpha_{k}^{\dag}\alpha_{k}\rangle$ and $\langle \beta_{k}^{\dag}\beta_{k}\rangle$ are given by the usual Bose-Einstein distribution:
\ba
\langle \alpha_{k}^{\dag}\alpha_{k}\rangle=\frac{1}{e^{(\varepsilon_{\alpha}(k)-\mu_\alpha)/k_B T}-1} \\ \nonumber
\langle \beta_{k}^{\dag}\beta_{k}\rangle=\frac{1}{e^{(\varepsilon_{\beta}(k)-\mu_\beta)/k_B T}-1} \nonumber
\ea
where $\mu_{\alpha,\beta}$ are the chemical potentials of the condensates of quasiparticles, $k_B$ is a Boltzmann constant and $T$ is e temperature of the condensate.

The chemical potentials can be obtained by fixing the average number of boson per site (filling). It is convenient to introduce the new variables $\mu=(\mu_{\alpha}+\mu_{\beta})/2$, $\delta=(\mu_{\alpha}-\mu_{\beta})/2$. The partition function  of the system is given by
\be
\textit{Z}=\prod_k[1-e^{-\beta(\varepsilon_{\alpha}(k)-\mu)}][1-e^{-\beta(\varepsilon_{\beta}(k)-\mu)}]
\ee
where $\beta=1/k_B T$. The free energy of the system can be calculated from the partition function
\be
F=-\frac{1}{N\beta}\ln{\textit{Z}}
\ee
Then the chemical potentials can be fixed solving  the following equations:
\be
N_{\alpha}+N_{\beta}=-\frac{\partial F}{\partial \delta}, \  \ N_{\alpha}-N_{\beta}=-\frac{\partial F}{\partial \mu}
\ee
where the $N_{\alpha,\beta}$ are the numbers of the quasiparticles of the type $\alpha$ and $\beta$ respectively.It is easy to show, that $N_{\alpha}+N_{\beta}=N_T$,where $N_T$ is a total number of the bosonic particles in the system.

\end{appendix}

\end{document}